\documentclass[submission,copyright,creativecommons]{eptcs}
\providecommand{\event}{GandALF 2020} 
\usepackage{underscore}           

\usepackage{amsthm,amsmath,amssymb}
\usepackage[utf8]{inputenc}
\usepackage[T1]{fontenc}
\usepackage[english]{babel}
\usepackage{mathtools}
\usepackage[caption=false]{subfig} 
\usepackage{enumitem} 
\usepackage{microtype}
\usepackage{algorithm}
\usepackage{algorithmic}
\usepackage{tikz}
\usepackage{sts} 

\newtheorem{proposition}{Proposition}[section]
\newtheorem{lemma}[proposition]{Lemma}

\theoremstyle{definition}
\newtheorem{definition}[proposition]{Definition}
\newtheorem{example}[proposition]{Example}
\newtheorem{remark}[proposition]{Remark}

\makeatletter
\renewcommand{\ALG@name}{Procedure}
\makeatother

\title{Decisiveness of Stochastic Systems\\ and its Application to Hybrid Models\thanks{Research supported by F.R.S.-FNRS under Grant n$^\circ$ F.4520.18 (ManySynth), and ENS Paris-Saclay visiting professorship (M.\ Randour, 2019). Pierre Vandenhove is an F.R.S-FNRS Research Fellow. Mickael Randour is an F.R.S-FNRS Research Associate.}}
\author{Patricia Bouyer
\institute{LSV, CNRS \& ENS Paris-Saclay, Universit{\'e} Paris-Saclay, France}
\and
Thomas Brihaye
\institute{UMONS -- Universit{\'e} de Mons, Belgium}
\and
Mickael Randour
\institute{F.R.S.-FNRS \& UMONS -- Universit{\'e} de Mons, Belgium}
\and
C{\'e}dric Rivi{\`e}re
\institute{UMONS -- Universit{\'e} de Mons, Belgium}
\and
Pierre Vandenhove
\institute{F.R.S.-FNRS \& UMONS -- Universit{\'e} de Mons, Belgium}
\institute{LSV, CNRS \& ENS Paris-Saclay, Universit{\'e} Paris-Saclay, France}
}

\def\titlerunning{Decisiveness of Stochastic Systems and its Application to Hybrid Models}
\def\authorrunning{Bouyer, Brihaye, Randour, Rivi{\`e}re \& Vandenhove}

\begin{document}

\maketitle

\begin{abstract}
In~\cite{ABM07}, Abdulla et al.\ introduced the concept of
decisiveness, an interesting tool for lifting good properties of
finite Markov chains to denumerable ones. Later~\cite{BBBC18}, this
concept was extended to more general stochastic transition systems (STSs), allowing the
design of various verification algorithms for large classes of (infinite)
STSs. We further improve the understanding and utility of decisiveness in two ways.

First, we provide a general criterion for proving the decisiveness of
general STSs. This criterion, which is very
natural but whose proof is rather technical, (strictly) generalizes
all known criteria from the literature.
Second, we focus on stochastic hybrid systems (SHSs), a
stochastic extension of hybrid systems. We establish the
decisiveness of a large class of SHSs and, under a few classical
hypotheses from mathematical logic, we show how to decide reachability
problems in this class, even though they are undecidable for
general SHSs. This provides a decidable stochastic extension of
o-minimal hybrid systems~\cite{BMRT04,Gen05,LPS00}.
\end{abstract}

\section{Introduction}
\subparagraph{Hybrid and stochastic models.}
Various kinds of mathematical models have been proposed to represent real-life
systems. In this article, we focus on models combining \emph{hybrid} and
\emph{stochastic} aspects. We outline the main features of these models to
motivate our approach.

The idea of \emph{hybrid systems} originates from
the urge to study systems subject to both \emph{discrete} and \emph{continuous}
phenomena, such as digital computer systems interacting with analog data.
These systems are transition systems with two kinds of transitions: continuous
transitions, where some continuous variables evolve over time
(e.g., according to a differential equation), and discrete transitions, where the system changes
modes and variables can be reset.
Much of the research about hybrid systems focuses on \emph{non-deterministic}
hybrid systems, i.e., systems modeling uncertainty by considering all possible
behaviors (e.g., different possibilities for discrete transitions
at a given time, arbitrary long time between discrete transitions).
A typical
question concerns the \emph{safety} of such systems---if a system can
reach an undesirable state, it is said to be \emph{unsafe}; if not, it is
\emph{safe}---such a
specification is called \textit{qualitative}. However, this is limiting for two
reasons. First, it does not take
into account that some behaviors are more likely to occur than others. Second,
risks cannot necessarily be avoided, and it is unrealistic to prevent
undesirable outcomes altogether. Therefore, we want to make probabilities an
integral part of our models,
in order to be able to \textit{quantify} the probability that
they behave according to the specification.
We thus consider the class of \emph{stochastic hybrid systems} (SHSs, for
short), hybrid systems in which a stochastic semantics replaces non-determinism.

\subparagraph{Goals.}
Our interest lies in the formal analysis of continuous-time SHSs, and more
specifically in \emph{reachability} questions, i.e., concerning the likelihood
that some set of states is reached. The
questions we seek to
answer are both of the qualitative kind (is some region of the state space
\emph{almost surely} reached, i.e., reached with probability $1$?) and of
the quantitative kind (what is the probability
that some states are eventually reached?). Such questions are
crucial, as verifying that a system works safely often reduces to verifying
that some undesirable state of the system is never reached (or with a
low probability), or that some desirable state is to be reached with high
probability~\cite{BK08}.
We want to give algorithms that decide, for an SHS $\calH$ and
reachability property $P$, whether $P$ is satisfied in $\calH$.
Such an endeavor faces multiple challenges; a first obvious one being that even
for rather restricted classes of non-deterministic hybrid systems, reachability
problems are undecidable~\cite{HKPV,HR00}. We want to define and
consider a \emph{class} of SHSs for which \emph{some} reachability problems are
decidable.

\subparagraph{Methods and contributions.}
Our methodology consists of two main steps. In a first step, we
follow the approach of Bertrand et al.~\cite{BBBC18}: we study general
\emph{stochastic transition systems} (STSs) through the
\emph{decisiveness} concept (Section~\ref{sec:sts}). The class of STSs
is a very versatile class of systems encompassing many well-known
families of stochastic systems, such as Markov chains,
generalized semi-Markov processes, stochastic timed automata, stochastic Petri
nets, and stochastic hybrid systems. Decisiveness was introduced
in~\cite{ABM07} to study Markov chains, and extended to STSs
in~\cite{BBBC18}. An STS is said to be \emph{decisive} with respect to
a set of states $B$ if executions of the system almost surely reach $B$ or a state from which $B$ is unreachable. Decisive
STSs benefit from useful properties that make possible the design of
verification algorithms related to reachability properties. Our
first contribution is to provide a \textbf{criterion to check the
 decisiveness of STSs} (Proposition~\ref{cor:mainDecCrit}), which
generalizes the decisiveness criteria from~\cite{ABM07,BBBC18}. This
generalization was mentioned as an open problem in~\cite{BBBC18}.

In a second step, we focus on \textit{stochastic hybrid systems}, which we introduce in Section~\ref{sec:stochHSDef}.
Reachability problems in hybrid systems are notoriously undecidable~\cite{HKPV,HR00}.
Our contributions regarding SHSs are split into two parts.

First, we aim to use the decisiveness idea to get closer to
the decidability frontier.
Albeit desirable, the decisiveness of a class of SHSs is not
sufficient to handle algorithmic questions about each SHS, as we need an
effective way to apprehend their uncountable state space. In this regard, an
often-used technique is to consider
a \emph{finite abstraction} of the system, that is, a finite partition of the
state space that preserves the properties to be verified (a well-known
example is the region graph for timed automata~\cite{AD94}).
To find such an abstraction of SHSs, we borrow ideas from~\cite{BMRT04,LPS00}:
we consider SHSs with \emph{strong resets} (Section~\ref{sec:decisiveSHS}), a
syntactic condition that decouples their continuous behavior from
their discrete behavior. We show that \textbf{SHSs with adequately placed
strong resets} (at least one per cycle of their discrete graph) $(i)$
\textbf{have a finite abstraction} (Proposition~\ref{prop:finiteAbs}), and
$(ii)$ \textbf{are decisive} (Proposition~\ref{prop:strongResetAreDec}), which
can be proved using our new criterion.

Second, in Section~\ref{sec:ominimal}, we show, with strong resets, how to effectively compute a finite abstraction and use
it to perform a reachability analysis of the original system.
We proceed by assuming that the components of our systems are
definable in an \emph{o-minimal structure}. The main difficulty here is that ``\emph{a satisfactory theory of measure and integration seems to
be lacking in the o-minimal context}''~\cite{BO04}; in an
o-minimal structure, the primitive function of a definable function
is in general not definable, which complicates
definability questions regarding probabilities. We therefore restrict
the possible probability distributions, using properties of
o-minimal structures to keep our class as large as possible.
When the \textbf{theory} of the structure is \textbf{decidable} (as is the case
for the ordered field of real numbers~\cite{Tar48}), the \textbf{reachability
problems are then decidable}.
This provides a stochastic extension to the theory of o-minimal hybrid systems~\cite{LPS00}.
We study \textbf{qualitative} and \textbf{quantitative} problems.
Due to space constraints, all the missing proofs and some technical details and additional results are omitted from this version and can be found in the full article~\cite{BBRRV20}.

\subparagraph{Related work.}
Our results combine previous work on stochastic systems and hybrid
systems. About stochastic systems, we build on the work of~\cite{ABM07,BBBC18}.
Our work also takes inspiration from research about \emph{stochastic timed
automata}, a subclass of stochastic hybrid systems which already combines
stochastic and timed aspects;
model checking of stochastic timed automata has been studied
in~\cite{BBBBG08,BBBM08,BBB+14} and
considered in the context of the decisiveness property in~\cite{Car17}.
Fundamental results about the decidability of the reachability problem for
hybrid systems can be found in~\cite{AD94,HKPV,HR00}.
The class of \emph{o-minimal hybrid systems}, of which we
introduce a stochastic extension, has been studied
in~\cite{BMRT04,Gen05,LPS00}.

The literature about SHSs often follows a more practical or numerical
approach. A first introduction to the model was provided in~\cite{HLS00}. An
extensive review of the underlying theory and of many applications of SHSs
is provided in books~\cite{CL07,Bu12}, and a review of different possible
semantics for this model is provided in~\cite{LP10}.
Applications of SHSs are numerous: a few examples are air traffic
management~\cite{PH09,PHLS00}, communication networks~\cite{He04}, biochemical
processes~\cite{LOQD17,SH10}.
The software tool \textsc{Uppaal}~\cite{DDL+12}
implements a model of SHS similar to the one studied in this article and uses
numerical methods to compute reachability probabilities, through numerical solving of differential equations and Monte Carlo simulation.
Reachability problems have also been considered in an alternative semantics
with \emph{discrete time}; a numerical approach is for instance provided
in~\cite{AKLP10,APLS08}.

\subparagraph{Notations.}
We write $\IR^+ = \{x\in\IR\mid x \ge 0\}$ for the set of non-negative real numbers.
Let $(\Omega,\Sigma)$ be a measurable space. We write
$\Dist(\Omega,\Sigma)$ (or $\Dist(\Omega)$ if there is no ambiguity) for
the set of \emph{probability} distributions over $(\Omega,\Sigma)$.
The complement of a set $A\in\Sigma$ is denoted by $\comp{A} =
\Omega\setminus A$.
For $A\in\Sigma$,
we say that two probability distributions $\mu, \nu \in \Dist(\Omega)$ are
\emph{(qualitatively) equivalent on $A$} if for
each $B\in\Sigma$, if $B\subseteq A$, $\mu(B)>0$ if and only if $\nu(B)>0$.

\section{Decisiveness of Stochastic Transition Systems}
\label{sec:sts}
In this section, we define \emph{stochastic transition systems} (STSs,
for short) as in~\cite{BBBC18}.
We then describe
the concept of \emph{decisiveness}, first defined in the specific case
of Markov chains~\cite{ABM07}, and then extended to STSs~\cite{BBBC18}.
Decisive stochastic systems benefit from ``nice'' properties making their
qualitative and quantitative analysis more accessible. The first
contribution of our work is a new decisiveness criterion
(Proposition~\ref{cor:mainDecCrit}), which generalizes existing criteria
from the literature~\cite{ABM07,BBBC18}. It is an intuitive criterion,
which was conjectured in~\cite{BBBC18} but could not be proved.
We finish with a brief section on the notion of \emph{abstraction}
between STSs, which will be useful to apply our results to stochastic
hybrid systems.

\subsection{Stochastic Transition Systems and Decisiveness\texorpdfstring{~\cite{BBBC18}}{}}

\begin{definition}[Stochastic transition system]
	A \emph{stochastic transition system} (STS) is a tuple
	$\calT =(S,\Sigma,\kappa)$ consisting of a measurable space of \emph{states}
	$(S,\Sigma)$, and a function $\kappa\colon S \times\Sigma \to [0,1]$ (sometimes called \emph{Markov kernel}) such that
	for every $s \in S$, $\kappa(s,\cdot)$ is a probability measure
	and for every $A \in \Sigma$, $\kappa(\cdot, A)$ is a
	measurable function.
\end{definition}

The second condition on $\kappa$ implies in particular that for a
measurable set $B\in\Sigma$, the set $\{s\in S\mid \kappa(s, B) > 0\}$ is measurable.

We interpret STSs as systems generating runs, with a probability
measure over these runs.
We fix an STS $\calT=(S,\Sigma,\kappa)$. From a state $s \in S$, a
probabilistic transition is performed according to distribution
$\kappa(s, \cdot)$, and the system resumes from one of the successor states;
this process generates random sequences of states. A \emph{run} of $\calT$ is an infinite sequence $s_0s_1s_2\ldots$ of states.
To formally provide a probabilistic semantics to STSs, we define a probability measure over the set of runs of $\calT$.
From an initial distribution $\mu\in\Dist(S)$, we can define in a classical way a unique probability measure $\Prob_{\mu}^\calT$ on the $\sigma$-algebra generated by all the \emph{cylinders}, using Carath{\'e}odory's extension theorem.

\subparagraph{Expressing properties of runs.}
We use standard \LTL notations~\cite{Pnu77} to express properties of runs of $\calT$.
If $B, B' \in \Sigma$, we write in particular $\F B$ (resp.\ $\F[\le n] B$, $B'\U B$, $B'\U[\le n] B$, $\GG B$, $\GG \F B$) for the set of runs that visit $B$ at some point (resp.\ visit $B$ in less than $n$ steps, stay in $B'$ until a first visit to $B$, stay in $B'$ until a first visit to $B$ in less than $n$ steps, always stay in $B$, visit $B$ infinitely often).
We are interested in two kinds of reachability problems.

\begin{definition}[Qualitative and quantitative reachability]
\label{def:problems}
	Let $B\in\Sigma$ be a measurable set of target states, and $\mu\in\Dist(S)$
	be an initial distribution. The \emph{qualitative reachability problems}
	consist in deciding whether $\Prob^\calT_\mu(\F B) = 1$, and whether
	$\Prob^\calT_\mu(\F B) = 0$. The
	\emph{quantitative reachability problem} consists in deciding, given
	$\epsilon, p\in\IQ$ with $\epsilon > 0$, whether $\abs{\Prob^\calT_\mu(\F
	B) - p} < \epsilon$.
\end{definition}

\subparagraph{Transforming probability distributions.}
Another useful way to reflect on STSs is as transformers of
probability distributions on $(S,\Sigma)$.

\begin{definition}[STS as a transformer]
	For $\mu \in\Dist(S)$, its \emph{transformation through $\calT$} is the
	probability distribution $\Omega_\calT(\mu)\in\Dist(S)$ defined for
	$A\in\Sigma$ by
	\[
	\Omega_\calT(\mu)(A) = \int_{s\in S} \kappa(s, A)\, \ud\mu(s)
	\puncteq{.}
	\]
\end{definition}

The value $\Omega_\calT(\mu)(A)$ is the probability to reach $A$
in one step, from the initial distribution $\mu$.

\subparagraph{Attractors.}
We will be particularly interested in the existence of \emph{attractors} for STSs.

\begin{definition}[Attractor]
	A set $A\in\Sigma$ is an \emph{attractor for $\calT$} if for every
	$\mu\in\Dist(S)$, $\Prob^{\calT}_{\mu}(\F A)=1$.
\end{definition}

While $S$ is always an attractor for $\calT$, we will later search for
attractors with more interesting properties.
The definition of \emph{attractor} actually implies a seemingly
stronger statement: an attractor is almost surely visited
\emph{infinitely often} from any initial
distribution~\cite[Lemma~19]{BBBC18}.

\subparagraph{Decisiveness.}
Before introducing decisiveness,
we give the definition of an \emph{avoid-set}: for
$B \in \Sigma$, its \emph{avoid-set} is written as
$\Btilde = \{s \in S \mid \Prob_{\delta_s}^{\calT}(\F B) = 0\}$ (where $\delta_s$ is the Dirac distribution at $s$). The
avoid-set $\Btilde$ corresponds to the set of states from which
runs almost surely stay out of $B$ \textit{ad infinitum}. One
can show that the set $\Btilde$ belongs to the $\sigma$-algebra
$\Sigma$~\cite[Lemma~14]{BBBC18}. We can now define the concept of
decisiveness as in~\cite{BBBC18}.

\begin{definition}[Decisiveness]
	Let $B\in\Sigma$ be a measurable set. We say that $\calT$ is \emph{decisive
	w.r.t.\ $B$} if for every $\mu\in\Dist(S)$,
	$\Prob^{\calT}_\mu(\F B \lor \F \Btilde) = 1$.
\end{definition}

Intuitively, the decisiveness property states that, almost surely,
either $B$ will eventually be visited, or states from which $B$ can no
longer be reached will eventually be visited.

\begin{example}[Random walk]
	We consider the STS $\calT$ (random walk) from Figure~\ref{fig:randomWalk}.
	We want to find out whether $\calT$ is decisive
	w.r.t.\ $B = \{0\}$. We assume that the initial distribution is given by
	$\delta_1$ (the Dirac distribution at $1$). By the theory on random walks,
	if $\frac{1}{2} < p < 1$, the walk will almost surely diverge to
	$\infty$. This entails that $\Prob^\calT_{\delta_1}(\F B) < 1$.
	Moreover, since $p < 1$, there is a path with
	positive probability from every state to $0$, so $\Btilde =	\emptyset$.
	Therefore,
	$
	\Prob^\calT_{\delta_1}(\F B \lor \F\Btilde) = \Prob^\calT_{\delta_1}(\F B)
	< 1\puncteq{,}
	$
	which means that $\calT$ is not decisive w.r.t.\ $B$.
	If $p\le\frac{1}{2}$, state $0$ is almost surely reached from any state, that is, for any initial distribution $\mu\in\Dist(S)$, we have that
	$\Prob^\calT_\mu(\F B) = 1$. Hence, in this case, STS
	$\calT$ is decisive w.r.t.\ $B$.
\end{example}

\begin{figure}[tbh]
	\centering
	\begin{tikzpicture}[scale=0.025, trim left]
	\node[font=\large] at (232, 34.018) {$\ldots$};
	\node[font=\large] at (16, 34.018) {$0$};
	\node[font=\large] at (88, 34.018) {$1$};
	\node[font=\large] at (160, 34.018) {$2$};
	\node[font=\large] at (52, 62.018) {$1$};
	\node[font=\large] at (124, 62.018) {$p$};
	\node[font=\large] at (196, 62.018) {$p$};
	\node[font=\large] at (196, 6.018) {$1-p$};
	\node[font=\large] at (124, 6.018) {$1 - p$};
	\node[font=\large] at (52, 6.018) {$1 - p$};
	\draw
	(16, 34.0175) circle[radius=16];
	\draw
	(88, 34.0175) circle[radius=16];
	\draw
	(160, 34.0175) circle[radius=16];
	\draw[->]
	(26.583, 46.0175)
	arc[start angle=-120.0133, end angle=-59.9867, x radius=50.8136, y
	radius=-50.8136];
	\draw[->]
	(98.583, 46.0175)
	arc[start angle=-120.0133, end angle=-59.9867, x radius=50.8136, y
	radius=-50.8136];
	\draw[->]
	(149.417, 22.0175)
	arc[start angle=59.9867, end angle=120.0133, x radius=50.8136, y
	radius=-50.8136];
	\draw[->]
	(77.417, 22.0175)
	arc[start angle=59.9867, end angle=120.0133, x radius=50.8136, y
	radius=-50.8136];
	\draw[<-]
	(170.583, 22.0175)
	arc[start angle=-120.0133, end angle=-60.0039, radius=50.8136];
	\draw[->]
	(170.583, 46.0175)
	arc[start angle=-120.0133, end angle=-59.9347, x radius=50.8136, y
	radius=-50.8136];
	\end{tikzpicture}
	\caption{STS $\calT$ representing a random walk on $\IN$.}
	\label{fig:randomWalk}
\end{figure}

A major interest of the decisiveness concept lies in the design of
simple procedures for the qualitative and quantitative analysis of stochastic
systems.
Indeed, as presented in~\cite{ABM07,BBBC18}, it allows,
under effectiveness hypotheses, to verify qualitative properties or to compute arbitrary close approximations of the
probabilities of various properties, like reachability, repeated reachability, and even $\omega$-regular properties. The reader is referred to~\cite[Sections~6 \&~7]{BBBC18} for more details, but we briefly recall the
approximation scheme for reachability properties in order to illustrate the usefulness of the decisiveness property. This scheme will be applied to a specific class of STSs in Section~\ref{sec:ominimal}.

Let $B \in \Sigma$ be a measurable set and $\mu \in \Dist(S)$ be an
initial distribution. To compute an approximation of
$\Prob_\mu^\calT(\F B)$, we define two sequences $(\pnYes)_{n\in\IN}$
and $(\pnNo)_{n\in\IN}$ such that for $n\in\IN$,
\[
\pnYes = \Prob_\mu^\calT(\F[\leq n] B)\ \text{and}\
\pnNo = \Prob_\mu^\calT(\comp{B} \U[\leq n]{\Btilde})\puncteq{.}
\]
These sequences are non-decreasing and converge respectively to
$\Prob_\mu^\calT(\F B)$ and $\Prob_\mu^\calT(\comp{B} \U \Btilde)$.
Observe moreover that for all $n \in \IN$, we have that $\pnYes \leq \Prob_\mu^\calT(\F B) \leq 1 - \pnNo$.

The main idea behind decisiveness of STSs lies in the following
property~\cite{ABM07,BBBC18}: if $\calT$ is decisive w.r.t.\ $B$, then
$\lim_{n \to \infty} \pnYes + \pnNo = 1$.
Therefore, for any given $\epsilon > 0$, for some $n$ sufficiently
large,
$\pnYes \leq \Prob_\mu^\calT(\F B) \leq \pnYes + \epsilon$.
In situations where $\pnYes$ and $\pnNo$ can be effectively
approximated arbitrarily closely and $\calT$ is decisive w.r.t.\
$B$, we can thus approximate
$\Prob_\mu^\calT(\F B)$ up to any desired error bound.

\subsection{A New Criterion for Decisiveness}
Our goal is to provide new sufficient conditions for the decisiveness of STSs.
To this end, we expose the following crucial technical lemma.

\begin{lemma} \label{lem:GBcGFA}
	Let $B \in \Sigma$, and $A \in \Sigma$.
	Suppose that there is $p > 0$ such that for all $\nu \in \Dist(A)$, we have
	$\Prob^\calT_{\nu}(\F B) \geq p$. Then for any
	$\mu \in\Dist(S)$,
	$\Prob^\calT_\mu(\GG \comp{B} \land \GG\F A) = 0$.
\end{lemma}

This result seems rather intuitive: if we visit $A$ infinitely often, and
after every passage through $A$ we have a probability bounded from
below to reach $B$, then the probability to stay in $\comp{B}$ forever is $0$.
An equivalent statement for Markov
chains has been used without proof in~\cite[Lemmas~3.4 \&~3.7]{ABM07}.
A weaker version of this statement is given as part of the proof of~\cite[Proposition~36]{BBBC18}, where it is said that this general case was not known to be true or false.
This weaker version assumes that there is a uniform upper bound $k$ such that for all $\nu\in\Dist(A)$, $\Prob^\calT_\nu(\F[\le k] B) \geq p$ to obtain a similar conclusion.
We have removed the need for this constraint.

A possible proof of this result consists of regarding a stochastic transition system as a stochastic process, and resorting to results from martingale theory.
More precisely, using \textit{L\'evy's zero-one law}, we obtain that infinite runs that never reach $B$ are the same (up to a set of probability $0$) as the infinite runs $s_0s_1\ldots$ for which the probability to reach $B$ given $s_0\ldots s_n$ converges to $0$ as $n$ grows to infinity.
Runs that visit $A$ infinitely often cannot both avoid $B$ and have a probability to visit $B$ that converges to $0$ (since for every visit to $A$, the probability to visit $B$ is bounded from below by $p > 0$).
Therefore, such runs will almost surely visit $B$.

We can now state our main contribution to decisiveness.

\begin{proposition}[Decisiveness criterion] \label{cor:mainDecCrit}
	Let $B\in\Sigma$ be a
	measurable set, and $A\in\Sigma$ be an attractor for $\calT$. We
	denote $A' = A\cap\comp{(\Btilde)}$ the set of states of $A$ from
	which $B$ is reachable with a positive probability. Assume that
	there exists $p > 0$ such that for all $\nu\in\Dist(A')$,
	$\Prob_\nu^\calT(\F B)\ge p$. Then $\calT$ is decisive w.r.t.\ $B$.
\end{proposition}

With probability $1$, every run visits attractor $A$ infinitely
often, but the hypotheses imply a dichotomy between runs.
Some runs will reach a state of $A$ from which $B$ is
almost surely non-reachable (i.e., in $\Btilde$). The other
runs will go infinitely often through states of $A$ such that the
probability of reaching $B$ is lower bounded by $p$ (i.e., in
$A'$), and will almost surely visit $B$ by Lemma~\ref{lem:GBcGFA}.
This almost-sure dichotomy between runs is required to show
decisiveness.

This criterion strictly generalizes those used in the
literature. The criterion in~\cite[Lemma~3.4]{ABM07} assumes the existence of a finite attractor; the
criteria in~\cite[Propositions~36 \&~37]{BBBC18} assume some finiteness
property in an abstraction (see next section), which we do not.
In~\cite[Lemma~3.7]{ABM07}, a similar kind of property as ours is required from
all the states of the STSs, not only from an attractor. We can actually prove the same result using a slightly more general notion of \emph{attractor} (see \cite[Proposition~11]{BBRRV20}).

\subsection{Abstractions of Stochastic Transitions Systems}
\label{sec:abstractions}
Decisiveness and abstractions are deeply intertwined concepts, so we
briefly recall this notion~\cite{BBBC18} and related properties. We
let $\calT_1 = (S_1,\Sigma_1,\kappa_1)$ and
$\calT_2 = (S_2,\Sigma_2,\kappa_2)$ be two STSs, and
$\alpha\colon (S_1,\Sigma_1) \to (S_2,\Sigma_2)$ be a measurable
function. We say that a set $B \in \Sigma_1$ is \emph{$\alpha$-closed}
if $B = \alpha^{-1}(\alpha(B))$. To mean that $B$ is
$\alpha$-closed, we also say that $\alpha$ is \emph{compatible} with $B$.
Following~\cite{Bog07,GBK16}, we
define a natural way to transfer measures from $(S_1,\Sigma_1)$ to
$(S_2,\Sigma_2)$ through $\alpha$: the \emph{pushforward of $\alpha$}
is the function $\alpha_\# \colon \Dist(S_1) \to \Dist(S_2)$ such that
$ \alpha _\#(\mu)(B) = \mu(\alpha^{-1}(B)) $ for every
$\mu \in\Dist(S_1)$ and for every $B \in \Sigma_2$.

\begin{definition}[$\alpha$-abstraction]
	STS $\calT_2$ is an \emph{$\alpha$-abstraction} of $\calT_1$ if for
	all $\mu \in \Dist(S_1)$, $\alpha_{\#}(\Omega_{\calT_1}(\mu))$ is
	qualitatively equivalent to $\Omega_{\calT_2}(\alpha_{\#}(\mu))$.
\end{definition}

Informally, the two STSs then have the same ``qualitative'' single steps.
Later, we may speak of \emph{abstraction} instead of
$\alpha$-abstraction if $\alpha$ is clear in the context.

The objective of the notion of abstraction is that by finding an
$\alpha$-abstraction $\calT_2$ which is somehow simpler than $\calT_1$
(for example, with a smaller state space), we should be able to use
$\calT_2$ (with initial distribution $\alpha_\#(\mu)$) to analyze some
properties of $\calT_1$ (with initial distribution $\mu$).
To do so, we need to know which properties are preserved through an $\alpha$-abstraction.
As a first observation, positive probability of reachability properties is preserved. Stronger conditions are required to study almost-sure reachability properties through $\alpha$-abstractions. We select a key property of abstractions about that matter which will be useful in the subsequent sections.

\begin{definition}[Sound $\alpha$-abstraction]
	We say that $\calT_2$ is a \emph{sound} $\alpha$-abstraction of $\calT_1$ if for all $B\in\Sigma_2$, $\Prob^{\calT_2}_{\alpha_\#(\mu)}(\F B) = 1$
	implies $\Prob^{\calT_1}_\mu(\F \alpha^{-1}(B)) = 1$.
\end{definition}

Sound abstractions preserve almost-sure reachability properties from $\calT_2$
to $\calT_1$. A major result from~\cite{BBBC18} is that if $\calT_1$ is decisive w.r.t.\ $B$ and $\calT_2$ is a sound $\alpha$-abstraction, then for all $\mu\in\Dist(S_1)$,
$\Prob_\mu^{\calT_1}(\F \alpha^{-1}(B)) = 1$ if and only if
$\Prob_{\alpha_{\#}(\mu)}^{\calT_2}(\F B) = 1$. Additional results about abstractions are provided in~\cite[Section~2.4]{BBRRV20}.

\section{Application to Stochastic Hybrid Systems}
\label{sec:stochHS}
We choose to restrict our attention to a stochastic extension of the
well-studied \emph{hybrid systems}.
A \emph{hybrid system} is a dynamical system combining discrete and continuous
transitions. It can be defined as a non-deterministic automaton with a finite
number of continuous variables, whose evolution is described via an infinite
transition system.
Hybrid systems have been widely studied since their
introduction in the 1990s (e.g.,~\cite{ACH+,Hen96}). They are
effectively used to model various time-dependent reactive systems; systems that
need to take into account both continuous factors (e.g., speed, heat, time,
distance) and discrete factors (e.g., events, instructions) are ubiquitous.

We define hybrid systems and give them a fully stochastic semantics, yielding the class of \emph{stochastic hybrid systems} (SHSs).
We then show that under classical definability hypotheses, we can obtain decidability results for qualitative and quantitative reachability problems in a class of SHSs (even though such problems are undecidable in full generality), using decisiveness (along with our new decisiveness criterion) as a key tool.
This provides a stochastic extension to the study of \emph{o-minimal hybrid systems}~\cite{BMRT04,Gen05,LPS00}.

\subsection{Hybrid Systems with Probabilistic Semantics} \label{sec:stochHSDef}
We proceed with the definition of a (non-deterministic) \emph{hybrid system}.

\begin{definition}[Hybrid system]
	A \emph{hybrid system} (HS) is a tuple
	$\calH=(L,X,\calA,E,\flow,\inv,\calG,\calR)$
	where:
	$L$ is a finite set of \emph{locations} (discrete states);
	$X=\{x_1,\ldots,x_n\}$ is a finite set of \emph{continuous variables};
	$\calA$ is a finite alphabet of \emph{events};
	$E \subseteq L \times \calA \times L$ is a finite set of \emph{edges};
	for each $\ell \in L$, $\flow(\ell)\colon \IR^n\times \IR^+ \to \IR^n$
	is a continuous function describing the \emph{dynamics} in location $\ell$;
	$\inv$ assigns to each location a subset of $\IR^{n}$ called
	\emph{invariant};
	$\calG$ assigns to each edge a subset of $\IR^{n}$ called \emph{guard};
	$\calR$ assigns to each edge $e$ and valuation $\ve\in\IR^n$ a subset
	$\calR(e)(\ve)$ of $\IR^n$ called \emph{reset}.
	For $\ell\in L$, $e\in E$, we usually denote $\flow(\ell)$ and $\calR(e)$
	by $\flow_\ell$ and $\calR_e$.
\end{definition}

We denote the number of variables $|X|$ by $n$.
Given a hybrid system $\calH$, we define $S_\calH = L\times\IR^n$ as the states
of $\calH$.
We distinguish two kinds of transitions between states:
\begin{itemize}
	\item there is a \emph{switch-transition}
	$(\ell,\ve)\xrightarrow{a}(\ell',\ve')$ if there exists
	$e = (\ell,a,\ell')\in E$
	such that $\ve\in \inv(\ell)\cap\calG(e)$, $\ve'\in \calR_e(\ve)\cap
	\inv(\ell')$;
	\item there is a \emph{delay-transition}
	$(\ell,\ve)\xrightarrow{\tau}(\ell,\ve')$ if there exists $\tau\in\IR^+$
	such that for all $0 \leq \tau' \leq \tau$, $\flow_\ell(\ve,\tau') \in
	\inv(\ell)$ and $\ve' = \flow_\ell(\ve,\tau)$.
\end{itemize}
Informally, a switch-transition $(\ell,\ve)\xrightarrow{a}(\ell',\ve')$ means
that an edge
$e=(\ell, a, \ell')$ can be taken without violating any constraint: the value
$\ve$ of the continuous variables is an element of the invariant $\inv(\ell)$
and of the guard $\calG(e)$,
and there is a possible reset $\ve'$ of the variables which is an element of
the invariant $\inv(\ell')$. A delay-transition
$(\ell,\ve)\xrightarrow{\tau}(\ell,\ve')$ means that some time $\tau$ elapses
without changing the discrete location of the system---the only constraint is
that all the values taken by the continuous variables during this time are in
the invariant $\inv(\ell)$.

Given $s = (\ell, \ve)\in L \times \IR^n$ a state of
the hybrid system, and $\tau \in \IR^+$, we denote by $s + \tau = (\ell,
\flow_\ell(\ve, \tau))$ the new state after some \emph{delay} $\tau$, without
changing the location.

We consider semantics given by \emph{mixed transitions}, i.e., transitions that
consist of a delay-transition (some time elapses)
followed by a switch-transition (an edge is taken and the location changes).
A mixed transition is denoted by $(\ell,\ve)\xrightarrow{\tau, a}(\ell', \ve')$
if and only if there exists $\ve''\in\IR^n$ such that
$(\ell,\ve)\xrightarrow{\tau}(\ell, \ve'')\xrightarrow{a}(\ell',\ve')$.

We usually assume that there is a bijection between the edges $E$ and the
alphabet of events $\calA$, and we omit mentioning this alphabet. If $e =
(\ell, a, \ell')\in E$, we can thus denote $\xrightarrow{e}$ (resp.\
$\xrightarrow{\tau, e}$) for switch-transitions (resp.\ mixed transitions)
instead of $\xrightarrow{a}$ (resp.\ $\xrightarrow{\tau, a}$).

\begin{example} \label{ex:pacman1}
	We provide in Figure~\ref{fig:pacman} an example of a hybrid system
	(first studied in~\cite{BBBBG08}). There are two continuous
	variables ($x$ and $y$) and five
	locations, each of them equipped with the same simple dynamics: $\dot{x} =
	\dot{y} = 1$ (i.e., $\gamma_\ell((x,y), \tau) = (x + \tau, y + \tau)$ for
	every location $\ell\in L$). Locations $\ell_2$ and $\ell_4$ have the same
	invariant,
	which is $\{(x,y)\mid y < 1\}$; the other invariants are simply $\IR^2$.
	Guards are written next to the edge to which they are
	related: for instance, $\calG(e_4) = \{(x,y)\mid y = 2\}$. The notation
	``$x\defeq 0$'' is used to denote a deterministic
	reset (in this case, the value of $x$ is reset to $0$ after taking
	the edge). For instance, $\calR_{e_1}(x, y)= \{(x, 0)\}$ (the value of
	$x$ is preserved and $y$ is reset to $0$).
	If nothing else is written next to an edge $e$, it means that there
	is no reset on $e$, i.e., that $\calR_{e}(\ve)= \{\ve\}$ for all $\ve\in\IR^n$.
	An example of mixed transitions of this system can be $(\ell_0, (0,0))
	\xrightarrow{0.4, e_0} (\ell_1, (0.4,0.4))
	\xrightarrow{0.6, e_1} (\ell_2, (1,0))
	\xrightarrow{0.2, e_2} (\ell_0, (0,0.2))
	\xrightarrow{1.5, e_3} (\ell_3, (1.5,1.7))$.
\end{example}

\begin{figure}[tbh]
	\centering
	\begin{tikzpicture}[scale = 0.0199, trim left]
	\node at (208.641, 52.623) {$\ell_0$};
	\node at (112.641, 52.623) {$\ell_1$};
	\node at (16.641, 52.623) {$\ell_2$};
	\node at (304.641, 52.622) {$\ell_3$};
	\node at (400.641, 52.622) {$\ell_4$};
	\node[font=\small] at (174.641, 104.623) {$x = 0$};
	\node[font=\small] at (174.641, 88.623) {$0\le y < 1$};
	\draw
	(208.6415, 52.6225) circle[radius=16];
	\draw
	(304.6415, 52.6225) circle[radius=16];
	\draw
	(400.6415, 52.6225) circle[radius=16];
	\node[font=\small] at (16.641, 76.623) {$y<1$};
	\node[font=\small] at (400.641, 76.623) {$y<1$};
	\draw[->]
	(224.6415, 52.6225)
	-- (288.6415, 52.6225);
	\draw[->]
	(320.6415, 52.6225)
	-- (384.6415, 52.6225);
	\node[font=\small] at (352.641, 62.623) {$y = 2$};
	\node[font=\small] at (352.641, 42.623) {$y \defeq 0$};
	\node[font=\tiny] at (324.641, 42.623) {$e_4$};
	\node[font=\small] at (256.641, 62.623) {$1 < y < 2$};
	\node[font=\tiny] at (228.641, 42.623) {$e_3$};
	\node[font=\small] at (304.641, 22.623) {$x > 2$};
	\node[font=\small] at (304.641, 2.623) {$x\defeq 0$};
	\draw[->]
	(400.6415, 36.6225)
	-- (400.6415, 12.6225)
	-- (224.6415, 12.6225)
	-- (208.6415, 36.6225);
	\draw[->]
	(208.6415, 108.6225)
	-- (208.6415, 68.6225);
	\node[font=\tiny] at (408.641, 28.623) {$e_5$};
	\draw
	(112.6415, 52.6225) circle[radius=16.4924];
	\draw
	(16.6415, 52.6225) circle[radius=16];
	\draw[->]
	(192.6415, 52.6225)
	-- (128.6415, 52.6225);
	\draw[->]
	(96.6415, 52.6225)
	-- (32.6415, 52.6225);
	\draw[->]
	(16.6415, 36.6225)
	-- (16.6415, 12.6225)
	-- (192.6415, 12.6225)
	-- (208.6415, 36.6225);
	\node[font=\small] at (64.641, 62.623) {$y = 1$};
	\node[font=\small] at (64.641, 42.623) {$y\defeq 0$};
	\node[font=\small] at (160.641, 62.623) {$y < 1$};
	\node[font=\small] at (112.641, 22.623) {$x > 1$};
	\node[font=\small] at (112.641, 2.623) {$x \defeq 0$};
	\node[font=\tiny] at (188.641, 42.623) {$e_0$};
	\node[font=\tiny] at (92.641, 42.623) {$e_1$};
	\node[font=\tiny] at (8.641, 28.623) {$e_2$};
	\end{tikzpicture}
	\caption{Example of a hybrid system with two continuous variables. Each
		location is equipped with the dynamics $\dot{x} = \dot{y} = 1$.}
	\label{fig:pacman}
\end{figure}

We give more vocabulary to refer to hybrid systems.
If there is a switch-transition $(\ell, \ve) \xrightarrow{e} (\ell', \ve')$, we
say that edge $e$ is \emph{enabled} at state $(\ell, \ve)$---this means that edge $e$ can be taken with no delay from state $(\ell, \ve)$.
Given a state $s = (\ell, \ve)$ and an edge $e = (\ell, a, \ell')$ of
$\calH$, we define $I(s, e) = \{\tau \in \IR^+ \mid s \xrightarrow{\tau, e}
s'\}$ as the set of delays after which edge $e$ is enabled from $s$, and $I(s)
= \bigcup_e I(s, e)$ as the set of delays after which any edge is
enabled from $s$.
For instance,
in the hybrid system from Figure~\ref{fig:pacman}, for $s =
(\ell_0,(0,0.2))$, we have $I(s,e_0) = \intervalco{0,0.8}$, and $I(s) =
\intervalco{0,1.8}\setminus\{0.8\}$.

We say that a state $s \in L \times \IR^n$ is \emph{non-blocking} if $I(s) \neq \emptyset$.
In the sequel, we only consider hybrid systems such that all states are
non-blocking, thereby justifying why considering solely mixed transitions is
doable---such a transition is available from any state.

We now replace the non-deterministic elements of hybrid systems with stochasticity.

\begin{definition}[Stochastic hybrid system]
	A \emph{stochastic hybrid system} (SHS) is
	defined as a tuple $\calH=(\calH',\delDist_L,\resDist_\calR,\edgDist)$, where:
	\begin{itemize}
		\item $\calH'=(L,X,\calA,E,\flow,\inv,\calG,\calR)$
		is a hybrid system,
		which is referred to as the \emph{underlying hybrid system of $\calH$}.
		We require guards, invariants and resets to be Borel sets.
		\item $\delDist_L \colon L \times \IR^n \to \Dist(\IR^+)$ associates to each state a probability distribution on the time delay in $\IR^+$ (equipped with the classical Borel $\sigma$-algebra) before leaving a location.
		Given $s \in L \times \IR^n$, the distribution
		$\delDist_L(s)$ will also be denoted by $\delDist_s$. We require that
		for every $s \in L \times \IR^n$, $\delDist_s(I(s))
		= 1$, i.e., the probability that an edge is enabled after a
		delay is $1$.
		\item $\resDist_\calR$ associates to each edge $e$ and valuation $\ve$
		a probability distribution on the set $\calR_e(\ve)\subseteq \IR^n$.
		Given $e\in E$ and $\ve\in\IR^n$, the
		distribution $\resDist_\calR(e)(\ve)$ will also be denoted by
		$\resDist_e(\ve)$.
		\item $\edgDist\colon L\times\IR^n \to \Dist(E)$
		assigns to each state of $\calH$ a probability distribution on the
		edges. We require that $\edgDist(s)(e) > 0$ if and only if edge
		$e$ is enabled at $s$. For $s\in L\times\IR$, we denote $\edgDist(s)$
		by $\edgDist_s$. This distribution is only defined for states at which
		an edge is enabled.
	\end{itemize}
\end{definition}

\begin{remark}
	The term ``stochastic hybrid system'' is used for a wide variety of
	stochastic extensions of hybrid systems throughout the literature.
	In this work, we consider stochastic delays, stochastic
	resets, a stochastic edge choice, and stochastic initial distributions.
	The way this probabilistic semantics is added on top of hybrid systems
	is very similar to how \emph{timed automata} are converted to
	\emph{stochastic timed automata} in~\cite{BBBBG08,BBBM08,BBB+14}.
	Although dynamics appear to be deterministic, the model is powerful enough to emulate stochastic dynamics by assuming that extra variables are
	solely used to influence the continuous flow of the other variables. These
	variables can be chosen stochastically in each location through the reset
	mechanism.
	This is for example sufficient to consider a
	stochastic extension of the \emph{rectangular automata}~\cite{HKPV}, whose
	variables evolve according to slopes inside an interval (such as
	$\dot{x}\in\intervalcc{1,4}$).
	Our model is very close to the one of the software tool
	\textsc{Uppaal}~\cite{DDL+12}.
	In Section~\ref{sec:ominimal}, we will identify the need to restrict the definition of some components of SHSs to ensure their definability; this is however not required for the results of Section~\ref{sec:decisiveSHS}.
\end{remark}

When referring to an SHS, we make in particular use of the same
terminology as for hybrid systems (e.g., runs, enabled edges, allowed delays
$I(\cdot)$) to describe its underlying hybrid system.

In order to apply the theory developed in Section~\ref{sec:sts}, we
give the semantics of an SHS $\calH$ as an STS $\calT_{\calH} =
(S_{\calH},\Sigma_{\calH},\kappa_{\calH})$. The set $S_{\calH}$ is the
set $L \times \IR^n$ of states of $\calH$, and $\Sigma_{\calH}$ is the
$\sigma$-algebra product between $2^L$ and the Borel $\sigma$-algebra
on $\IR^n$. To define $\kappa_\calH$, we first explain briefly the
role of each probability distribution in the definition of SHS.
Starting from a state $s = (\ell, \ve)$, a \emph{delay} $\tau$ is chosen
randomly, according to the distribution $\delDist_s$. From state
$s+\tau = (\ell,\ve')$, an edge $e=(\ell,a,\ell')$ (enabled in
$s+\tau$) is selected, following distribution $\edgDist_{s + \tau}$
(such an edge is almost surely available, as $\delDist_s(I(s)) = 1$ by
hypothesis). The next state will be in location $\ell'$, and the
values of the continuous variables are stochastically reset according
to the distribution $\resDist_e(\ve')$. We can thus define
$\kappa_{\calH}$ as follows: for $s = (\ell, \ve) \in S_\calH$, $B \in
\Sigma_\calH$,
\[
\kappa_\calH(s,B)
= \int_{\tau \in \IR^+}
\sum_{e = (\ell,a,\ell')\in E} \left(
\edgDist_{s+\tau}(e) \cdot
\int_{\ve'' \in \IR^n}
\ind_B(\ell',\ve'') \ud (\resDist_e(\flow_\ell(\ve, \tau)))(\ve'')
\right) \ud \delDist_s(\tau)
\]
where $\ind_B$ is the characteristic function of $B$.
It gives the probability to hit set $B \subseteq S_{\calH}$ from state $s$ in one step (representing a mixed transition).
The function $\kappa_{\calH}(s, \cdot)$ defines a probability distribution for all $s\in S_\calH$.

\begin{definition}[STS induced by an SHS]
	For an SHS $\calH$, we define $\calT_\calH = (S_\calH, \Sigma_\calH,
	\kappa_\calH)$ as the \emph{STS induced by $\calH$}.
\end{definition}

Thanks to the stochasticity of our models, we can reason about
both \emph{qualitative} and \emph{quantitative} reachability problems, as
defined in Definition~\ref{def:problems}.

\subparagraph{Undecidability.}
We provide a proof that qualitative and quantitative reachability problems for SHSs with ``simple'' features are undecidable. The undecidability in itself is not surprising, as some of the undecidability proofs from the literature about non-deterministic hybrid systems~\cite{HKPV,HR00} can be translated directly to our stochastic setting. Our goal with this new proof is to get as close as possible to the class that we will later (in Section~\ref{sec:ominimal}) show to be decidable, in order to outline as well as possible an undecidability frontier.

We prove undecidability even when constrained to purely
continuous distributions on time delays from any state, and very simple guards,
resets, and dynamics (that can be defined in simple mathematical structures).
This requires a distinct proof from~\cite{HKPV}.
The result from~\cite{HR00} is close to the one we want to
achieve, and we take inspiration from its proof.
The proof consists of reducing the \emph{halting problem for
two-counter machines} to deciding whether a measurable
set in an SHS is reached with probability $1$.

\begin{proposition}[Undecidability of SHSs]
	The qualitative reachability problems and the approximate
	quantitative reachability problem are undecidable for stochastic
	hybrid systems with purely continuous distributions on time delays, guards
	that are linear comparisons of variables and constants, and using
	positive integer slopes for the flow of the continuous variables. The
	approximate quantitative problem is moreover undecidable for any fixed
	precision $\epsilon < \frac{1}{2}$.
\end{proposition}

Although the proof is centered on showing the undecidability
of qualitative reachability problems, we get as a by-product the undecidability
of the approximation. Indeed, as the systems used throughout the proof reach
a target set $B$ with a probability that is either $0$ or $1$, the ability to
approximate $\Prob^{\calT_\calH}_\mu(\F B)$ with $\epsilon < \frac{1}{2}$ would
be sufficient to solve the qualitative problem. As the proof shows that these
qualitative problems are undecidable, we obtain that the approximate
quantitative problem is undecidable as well.
Our result also implies that deciding whether a state lies in $\Btilde$ is undecidable.

\subsection{The Cycle-Reset Hypothesis}
\label{sec:decisiveSHS}
The literature about non-deterministic hybrid systems suggests that to obtain
subclasses for which
the reachability problem becomes decidable, one must set sharp restrictions on
the continuous flow of the variables and/or on the discrete transitions (via
the \emph{reset} mechanism).
In this decidable spectrum lie for instance
the \emph{rectangular initialized automata}, which are quite permissive toward
the continuous evolution of the variables, but need strong hypotheses about
the discrete transitions~\cite{HKPV}.
Our approach lies at one end of this spectrum: we restrict the
discrete behavior by considering \emph{strong resets}, i.e., resets that forget
about the previous values of the variables, decoupling the discrete behavior
from the continuous behavior. We show that one strong reset per
cycle of the graph is sufficient to obtain valuable results, and we name this
property \emph{\cyclereset{}}.
This point of view has already been studied in~\cite{BMRT04,Gen05,LPS00} for
non-deterministic hybrid systems, and in~\cite{BBBM08} for stochastic timed automata.
Note that previous work about finite abstractions of non-deterministic hybrid systems (called \emph{time-abstraction bisimulations}) does not extend in general to ``stochastic'' abstractions (as defined in Section~\ref{sec:abstractions}), as there may be transitions that have probability $0$ to happen.

\begin{definition}[Strong reset, cycle-reset SHS]
	Given $\calH$ a stochastic hybrid system and $e$ an edge of $\calH$, we
	say that $e$ has a \emph{strong reset} (or is \emph{strongly reset}) if there exist $\calR_e^* \subseteq \IR^n$ and $\resDist_e^* \in \Dist(\calR_e^*)$ such that for all $\ve \in \calG(e)$, $\calR_e(\ve) = \calR_e^*$ and $\resDist_e(\ve) = \resDist_e^*$.
	We say that an SHS $\calH$ is \emph{\cyclereset{}} if for every simple cycle of $\calH$, there exists a strongly reset edge.
\end{definition}

If an edge $e$ is strongly reset, it stochastically resets all the continuous
variables when it is taken, and the stochastic reset does not depend on their
values.
We show two independent and very convenient results of \cyclereset{} SHSs: such
SHSs are \emph{decisive w.r.t.\ any measurable set} (the proof of this
statement relies on the decisiveness criterion from
Proposition~\ref{cor:mainDecCrit}), and \emph{admit a finite
abstraction}.

\subparagraph{Decisiveness of \cyclereset{} SHSs.}
We motivate this section with an example of a simple non-decisive SHS, which we
will use to show that our decisiveness result is tight.

\begin{example} \label{ex:stochPacman}
	We add a stochastic layer to the hybrid system of
	Example~\ref{ex:pacman1}, pictured in Figure~\ref{fig:pacman}. The
	distributions on the time delays in locations $\ell_0$, $\ell_2$ and
	$\ell_4$
	are uniform distributions on the interval of allowed delays. For instance,
	at state $s = (\ell_0, (x, y))$, the distribution $\delDist_s$ follows a
	uniform
	distribution on $\intervalco{0, 2 - y}$. In locations $\ell_1$ and $\ell_3$, the
	distributions on the delays are Dirac distributions. Reset and edge distributions are simply modeled as Dirac distributions.
	It is proved in~\cite[Section~6.2.2]{BBB+14} that this SHS is not decisive
	w.r.t.\ $B = \{\ell_2\}\times\IR^2$. The proof is quite technical and we do not recall it here; intuitively, at each passage through location $\ell_0$, the value of $y$ increases but stays bounded from above by $1$,
	which decreases the probability to take edge $e_0$ (and thus reach $B$); this decrease is too fast to ensure that $B$ is almost surely reached.
\end{example}

\begin{proposition} \label{prop:strongResetAreDec}
	Every \cyclereset{} SHS is decisive w.r.t.\ any measurable set.
\end{proposition}

Placing (at least) one strong reset per simple cycle is an easy syntactic way
to guarantee that almost surely, infinitely many strong resets are
performed, which is the actual sufficient property used in the proof. As
there are only finitely many edges, we can find a probability lower
bound $p$ on the probability to reach $B$ after any strong reset, as
required in the criterion of
Proposition~\ref{cor:mainDecCrit}. Notice that as shown in
Example~\ref{ex:stochPacman}, having independent flows for each
variable and resetting each variable once in each cycle is not
sufficient to obtain decisiveness; variables need to be reset \emph{on
the same discrete transition} in each cycle.

\subparagraph{Existence of a finite abstraction.}
We show that \cyclereset{} SHSs admit a finite $\alpha$-abstraction.
We first give a simple example showing that without one strong reset per cycle,
some simple systems do not admit a finite $\alpha$-abstraction compatible with
the locations.

\begin{example}
	Consider the SHS of Figure~\ref{fig:noFiniteAbs}. The
	self-loop edge of $\ell_0$ is the only edge not being strongly reset.
	We assume that we want to have an abstraction compatible with
	$s^*\defeq\{\ell_1\} \times\IR$.
	In order to obtain an abstraction, we must split $\{\ell_0\}\times\IR^n$ in
	$s_0\defeq\{\ell_0\}\times\intervaloo{-\infty,0}$
	and $\{\ell_0\}\times\intervalco{0,+\infty}$, as all the states of $s_0$
	can reach $s^*$ with a
	positive probability in one step, but none of the states of
	$\{\ell_0\}\times\intervalco{0,+\infty}$ can. Then,
	$\{\ell_0\}\times\intervalco{0,+\infty}$ must also be split into
	$s_1\defeq\{\ell_0\}\times\intervalco{0,1}$ and
	$\{\ell_0\}\times\intervalco{1,+\infty}$ because the states of $s_1$ can
	all reach $s_0$ with a positive probability in one step, but none of the
	states of $\{\ell_0\}\times\intervalco{1,+\infty}$ can. By iterating this
	argument, we find that the smallest $\alpha$-abstraction compatible with
	$\{\ell_1\}\times\IR$ is countably infinite, and the partition that it induces is
	composed of $s^*$, $s_0$ and $s_i = \{\ell_0\}\times\intervalco{i-1, i}$
	for $i\ge 1$. The underlying hybrid system actually belongs to the class of
	\emph{updatable timed automata}~\cite{BDFP04}, and the abstraction almost
	coincides with the \emph{region graph} of the automaton.
	The reasoning we have used is very close to the classical bisimulation algorithm for non-deterministic systems, and is explained in more detail in~\cite[Section~2.4]{BBRRV20}.
\end{example}

\begin{figure}[tbh]
	\centering
	\begin{tikzpicture}[scale=0.024, trim left]
	\node at (127.182, 53.221) {$x < 0$};
	\draw
	(63.182, 41.2205) circle[radius=24];
	\draw
	(191.182, 41.2205) circle[radius=24];
	\draw[->]
	(87.182, 41.2205)
	-- (167.182, 41.2205);
	\node at (127.182, 29.221) {$x \defeq 0$};
	\node at (63.182, 33.221) {$\dot{x} = 1$};
	\draw[->]
	(46.211, 58.1915)
	.. controls (31.182, 73.2205) and (23.182, 65.2205) .. (19.182, 58.5538)
	.. controls (15.182, 51.8872) and (15.182, 46.5538) .. (15.182, 41.2205)
	.. controls (15.182, 35.8872) and (15.182, 30.5538) .. (19.182, 23.8872)
	.. controls (23.182, 17.2205) and (31.182, 9.2205) .. (46.211, 24.2495);
	\draw[->]
	(208.153, 58.1915)
	.. controls (223.182, 73.2205) and (231.182, 65.2205) .. (235.182,
	58.5538)
	.. controls (239.182, 51.8872) and (239.182, 46.5538) .. (239.182,
	41.2205)
	.. controls (239.182, 35.8872) and (239.182, 30.5538) .. (235.182,
	23.8872)
	.. controls (231.182, 17.2205) and (223.182, 9.2205) .. (208.153,
	24.2495);
	\node at (31.182, 77.221) {$x\ge 0$};
	\node at (31.182, 5.221) {$x \defeq x - 1$};
	\node at (63.182, 49.221) {$\ell_0$};
	\node at (191.182, 49.221) {$\ell_1$};
	\node at (191.182, 33.221) {$\dot{x} = 1$};
	\node at (223.182, 77.221) {$x\in\mathbb{R}$};
	\node at (223.182, 5.221) {$x\defeq 0$};
	\end{tikzpicture}
	\caption{The time delays are given by exponential
		distributions from any state; resets are Dirac distributions.
		The smallest abstraction compatible with $\{\ell_1\}\times\IR$ is
		countably infinite.}
	\label{fig:noFiniteAbs}
\end{figure}

The \cyclereset{} assumption is sufficient to guarantee the existence of a finite abstraction, as formulated in the next proposition.

\begin{proposition} \label{prop:finiteAbs}
	Let $\calH$ be an SHS, and $B\in\Sigma_\calH$. If $\calH$ is \cyclereset,
	it has a finite and sound abstraction compatible with $B$ and with the
	locations.
\end{proposition}

The proof consists of showing that a stochastic adaptation of the classical \emph{bisimulation algorithm} terminates under the \cyclereset{} assumption. Combined with our decisiveness result, we can even show that this abstraction is sound.

\subsection{Reachability Analysis in Cycle-Reset Stochastic Hybrid Systems}
\label{sec:ominimal}
Our goal in this section is to perform a reachability analysis
of \cyclereset{} SHSs.
A first hurdle to circumvent is that arbitrary SHSs are in general difficult to apprehend algorithmically: for instance, the continuous
evolution of their variables may be defined by solutions of systems of
differential equations, which we do not know how to solve in general.
To make the problem more accessible, we follow the approach
of~\cite{BMRT04,LPS00} for non-deterministic hybrid systems by assuming that
some key components of our systems are definable in a mathematical structure.
We identify a large class of SHSs for which the finite abstraction from
Section~\ref{sec:decisiveSHS} is computable,
which makes \emph{qualitative} reachability problems decidable; namely, \emph{\cyclereset{} o-minimal SHSs defined in a decidable theory}.
We then identify sufficient hypotheses for
the \emph{approximate quantitative} problem to be decidable, in the form of a
finite set of probabilities that have to be approximately computable.

\subparagraph{Qualitative analysis.}
We assume that the reader is familiar with basic model-theoretic and logical terms---the reader is referred to~\cite{Ho97} for an introduction to the main concepts. In what follows, by \emph{definable}, we mean definable without parameters.

\begin{definition}[Definable SHS]
	Given a structure $\calM$, an SHS $\calH$
	is said to be \emph{defined in $\calM$} if
	for every location $\ell\in L$, $\flow_\ell$ is a function definable in
	$\calM$ and $\inv(\ell)$ is a set definable in $\calM$, and
	for every edge $e\in E$, the set $\calG(e)$ is definable and there exists a
	first-order formula $\resetFormula_e(\vect{x},\vect{y})$ such that
	$\ve'\in\calR_e(\ve)$ if and only if $\resetFormula_e(\ve,\ve')$ is true.
\end{definition}

Note that we require that the \emph{flow} of the dynamical
system in each location is definable in $\calM$, and not that it is the
solution to a definable system of differential equations.

Our goal is to prove that the class of \emph{\cyclereset{} o-minimal SHSs}, i.e., \cyclereset{} SHSs defined in an \emph{o-minimal structure} (introduced in~\cite{vdD84,PS86}), with additional assumptions on the probability distributions, is decidable. We will use that the finite sound abstraction from Proposition~\ref{prop:finiteAbs} is then computable, and that decisiveness (Proposition~\ref{prop:strongResetAreDec}) gives us a strong link between reachability properties of the abstractions and the original systems.

\begin{definition}[O-minimality]
	A totally ordered structure $\calM = \langle M, <, \ldots\rangle$ is
	\emph{o-minimal} if every definable subset of $M$
	is a finite union of points and open intervals (possibly unbounded).
\end{definition}

In other words, the
definable subsets of $M$ are exactly the ones that are definable with
parameters in $\langle M, < \rangle$. Some well-known structures are
o-minimal, such as
the ordered additive group of rationals $\langle\IQ,<,+,0\rangle$,
the ordered additive group of reals $\Rlin = \langle\IR,<,+,0,1\rangle$, the ordered field of reals $\Ralg =
\langle\IR,<,+,\cdot,0,1\rangle$, the ordered field of reals with the
exponential function $\Rexp = \langle\IR,<,+,\cdot,0,1,e^x
\rangle$~\cite{Wi96}.
There is no general result about the decidability of the theories of o-minimal
structures. A well-known case is the
Tarski-Seidenberg theorem~\cite{Tar48}, which
asserts that there exists a
quantifier-elimination algorithm for sentences in the first-order language of
real closed fields. This result
implies the decidability of the theory of $\Ralg$.
However, it is not known whether the theory of $\Rexp$ is decidable. Its decidability is implied by \emph{Schanuel's	conjecture}, a famous unsolved problem in transcendental number theory~\cite{MW}.
In this work, we define an \emph{o-minimal SHS} as an SHS defined in an o-minimal structure.

The o-minimality of $\calM$ implies that definable subsets of $M^n$ have a
very ``nice'' structure, described notably by the
\emph{cell decomposition theorem}~\cite{KPS86}.
In particular, every subset of $\IR^n$ definable in an o-minimal structure
belongs to the $\sigma$-algebra of Borel sets of
$\IR^n$~\cite[Proposition~1.1]{Kai12}.
Moreover, to help with issues related to the definability of probabilities, we will use the property that in o-minimal structures, definable sets with positive Lebesgue measure coincide with definable sets with non-empty interior~\cite[Remark~2.1]{Kai12}. This is very helpful, as the property of having an empty interior is definable.

\begin{remark}
The definition of \emph{o-minimal (non-deterministic) hybrid system} in the
literature usually assumes that all edges are strongly reset. \emph{O-minimal
hybrid systems} were first introduced in~\cite{LPS00}, and further studied
notably in~\cite{BMRT04}. The strong reset hypothesis was relaxed
in~\cite{Gen05} to ``one strong reset per cycle''.
\end{remark}

Let $\calM = \langle\IR,<,+,\ldots\rangle$ be an o-minimal structure whose
theory is decidable, such as $\Ralg$. Let
$\calH=(\calH',\delDist_L,\resDist_\calR,\edgDist)$ be a \cyclereset
o-minimal SHS defined in $\calM$.
Let $\mu\in\Dist(S_\calH)$ be an initial distribution. We make the following
assumptions, which we denote by \hyp:
\begin{itemize}
	\item for all $s = (\ell,\ve)\in L\times\IR^n$, if $I(s)$ is finite,
	$\delDist_s$ is equivalent to the uniform discrete distribution on $I(s)$;
	if $I(s)$ is infinite, $\delDist_s$ is equivalent to the Lebesgue
	measure on $I(s)$;
	\item the initial distribution $\mu$ is either equivalent to the discrete measure on some finite definable support $D$, or equivalent to the Lebesgue measure on a definable support $D$;
	\item for $e\in E$, $\ve\in\IR^n$, we ask that $\calR_e(\ve)$ is
	finite or has positive Lebesgue measure; $\resDist_e(\ve)$
	is resp.\ either equivalent to the discrete measure
	or the Lebesgue measure on $\calR_e(\ve)$.
\end{itemize}

The first requirement was already a standard assumption in the case of stochastic timed automata~\cite{BBBBG08,BBBM08,BBB+14}.
Hypothesis \hyp is easily satisfied: for instance, exponential
distributions (resp.\ uniform distributions on $\intervalcc{a,b}$) are
equivalent to the Lebesgue measure on $\IR^+$ (resp.\ $\intervalcc{a,b}$).
We summarize what these ideas entail in the next proposition.

\begin{proposition}
	Let $\calH$ be a \cyclereset o-minimal SHS defined in a structure
	whose theory is decidable. Let $B\in\Sigma_\calH$ be definable and
	$\mu\in\Dist(S_\calH)$. We assume that
	assumption \emph{\hyp} holds. Then one can decide whether
	$\Prob^{\calT_\calH}_\mu(\F B) = 1$ and whether $\Prob^{\calT_\calH}_\mu(\F
	B) = 0$.
\end{proposition}

In particular, we can decide the qualitative reachability problems for \cyclereset SHSs defined in $\Ralg$ and satisfying \hyp. Assuming Schanuel's
conjecture~\cite{MW}, we could extend this result to $\Rexp$.

\subparagraph{Approximate quantitative analysis.}
Under strengthened numerical hypotheses, we
can solve the approximate quantitative reachability problem in \cyclereset{} SHSs.
Let $\calH$ be a \cyclereset{} SHS, $B\in\Sigma_\calH$ and
$\mu\in\Dist(S_\calH)$. Our goal is to apply the approximation scheme
described in~\cite{BBBC18} in order to approximate
$\Prob^{\calT_\calH}_\mu(\F B)$. To do so, we require
that $\calT_\calH$ is decisive w.r.t.\ $B$, which is implied by the
\cyclereset{} hypothesis (Proposition~\ref{prop:strongResetAreDec}),
and the ability to compute for all $m\in\IN$, an arbitrarily close
approximation of
\[\pmYes = \Prob_\mu^\calT(\F[\leq m] B) \text{ and } \pmNo = \Prob_\mu^\calT(\comp{B} \U[\leq m]{\Btilde}).\]
Using the \cyclereset{} hypothesis, we can express these
probabilities as sums and products of a \emph{finite} number of probabilities of paths with bounded length
$b\in\IN$, where $b$ is the length of the longest path without
encountering a strong reset.
Details on this computation are available in~\cite[Section~5.2]{BBRRV20}.

\section{Conclusion}
\subparagraph{Summary.}
We presented in Section~\ref{sec:sts} how to solve reachability 
problems in stochastic transition systems via the \emph{decisiveness} 
notion, introduced in~\cite{ABM07,BBBC18}. We notably solved in 
Lemma~\ref{lem:GBcGFA} a technical question that was open in~\cite{BBBC18},
which helped formulate a general sufficient condition for decisiveness in Proposition~\ref{cor:mainDecCrit}, encompassing known decisiveness criteria from the literature.

In Section~\ref{sec:stochHS}, we considered \emph{stochastic hybrid systems} (SHSs).
We showed that SHSs with one \emph{strong reset} per cycle (\emph{\cyclereset{}}) are decisive (using our new decisiveness criterion), and admit a finite abstraction.
We then identified assumptions, using ideas from logic, leading to the effective computability of this abstraction, and to the decidability of qualitative and quantitative reachability problems.
These assumptions pertain to the definability of the different components of 
the SHSs in a decidable mathematical structure.
Combined with the previous decisiveness results, this abstraction can be used to decide qualitative reachability problems.
In particular, these results apply to \cyclereset{} SHSs defined in $\Ralg = \langle\IR,<,+,\cdot,0,1\rangle$ with well-behaved distributions.

\subparagraph{Possible extensions and future work.}
We identify some possible extensions of our results.

A first direction of study is to find other classes of decisive
stochastic systems that can be encompassed by our decisiveness
criterion (Proposition~\ref{cor:mainDecCrit}). In that respect, a
good candidate is the class of \emph{stochastic regenerative Petri
nets}~\cite{HPRV-peva12,PHV16}. An application of decisiveness
results to \emph{stochastic Petri nets} was briefly discussed
in~\cite[Section~8.3]{BBBC18}, but under severe constraints; we may
be able to relax part of these constraints with the generalized
criterion.

In Section~\ref{sec:ominimal}, we circumvent the issue of the definability of measures by using a specific property of the o-minimal structures (namely, that the Lebesgue measure of a definable set is positive if and only if the interior of that set is non-empty).
However, more powerful results exist about the compatibility of o-minimal structures and measure theory~\cite{Kai12}: some o-minimal structures are closed under integration with respect to a given measure (then called \emph{tame} measure). 
This consideration may help extend our results to a larger class that is less restrictive w.r.t.\ probability distributions, or in which actual probabilities may be definable.

\subparagraph{Acknowledgments.}
We would like to thank the anonymous reviewers for their valuable advice, which notably helped simplify the proof of Lemma~\ref{lem:GBcGFA}.

\bibliographystyle{eptcs}
\bibliography{stochomin.bib}

\end{document}